\documentclass[11pt]{article}
\usepackage{amsmath,amsfonts,amssymb,amsthm,setspace,tensor,physics}
\usepackage[utf8]{inputenc}
\usepackage{slashed}
\usepackage[nosort]{cite}
\usepackage[table]{xcolor}
\usepackage{dsfont}
\usepackage{mathrsfs}
\usepackage{upgreek}
\usepackage{graphicx}
\usepackage{float}
\usepackage{longtable}
\usepackage{lipsum}
\usepackage{verbatim}
\usepackage{leftidx}
\usepackage{bm}
\usepackage{marginnote}
\usepackage[top=2.8cm,bottom=2.8cm,left=2.0cm,right=2.0cm]{geometry}
\usepackage[colorlinks=true,linkcolor=black,citecolor=black,urlcolor=blue,filecolor=black]{hyperref}

\numberwithin{equation}{section}
\allowdisplaybreaks

\begin{document}

\vspace{30pt}
\begin{center}
{\Large\sc Coherent States of Non-Null Torus Knots}
\vspace{-5pt}
\par\noindent\rule{325pt}{0.5pt}

\vskip 1cm
{\sc 
Gabriel Canadas da Silva and Ion Vasile Vancea
}

\vspace{10pt}
{\it 
Department of Physics,\\
Federal Rural University of Rio de Janeiro,\\
Cx. Postal 23851, BR 465 Km 7, 23890-000 Serop\'{e}dica - RJ, Brazil}

\vspace{4pt}
{\tt\small \href{mailto: ggcanadas@gmail.com}{ggcanadas@gmail.com}}
\hspace{0.5em}
{\tt\small \href{mailto:ionvancea@ufrrj.br}{ionvancea@ufrrj.br}}

\vspace{30pt} {\sc\large Abstract}
\end{center}

We construct coherent states for the quantized electromagnetic field that correspond to the classical non-null torus knot solutions of Maxwell's equations in vacuum. We derive the displacement operators 
from the general relation between classical fields and coherent state amplitudes and verify the defining properties of coherent states through direct computation. We determine the observables of the model: field expectation values, energy density, Poynting vector, helicity, photon number, quadrature uncertainties, and correlation functions, and calculate their expectation values in the knotted coherent states in terms of the integer parameters 
$(n,m,l,s)$ of the classical solutions. As an example, we particularize the construction in the case of the Hopfion coherent state. These results establish the quantum-classical correspondence for this type of vacuum topological electromagnetic  systems.

\vspace{0.5cm} 


\newpage

%----------------------------------------------------------------
%----------------------------------------------------------------
\section{Introduction}
\label{sec:intro}

Coherent states play an important role in quantum electrodynamics and its applications, as the quantum states that most closely resemble classical electromagnetic fields \cite{Glauber1963, Sudarshan1963, Klauder1985}. They have several remarkable properties that make them helpful for understanding the quantum-classical correspondence: they are eigenstates of the annihilation operator, they saturate the Heisenberg uncertainty relation, they maintain their shape under time evolution, their expectation values follow classical equations of motion, and they exhibit Poissonian photon statistics characteristic of ideal laser light \cite{Gazeau2018}.

In recent years, there has been a growing interest in topological structures within classical electromagnetic fields. An important class of topological electromagnetic fields that has attracted attention recently was introduced by Ra\~{n}ada \cite{Ranada1989, Ranada1990}, based on previous observations by Trautman on the relation between Hopf fibrations and electromagnetic fields \cite{Trautman1977}. The Ra\~{n}ada-Hopf fields are solutions to vacuum Maxwell's equations whose field lines are linked circles forming Hopf fibrations. Their physical and mathematical properties have been studied in several works \cite{Ranada1992,Ranada1995,Ranada1997,Ranada1998,Ranada2003,
Arrayas2011,Arrayas2012a,Arrayas2017b,Arrayas2017c,Ranada2017a,Alves2017}. Arrayás and Trueba \cite{Arrayas2015} generalized the Ra\~{n}ada-Hopf idea to construct families of non-null torus knot solutions, where field lines are torus knots characterized by integers $(n,m,l,s)$.  These solutions exhibit non-trivial topology manifest in quantized helicities proportional to the linking numbers $nm$ and $ls$. 
Recent work has explored the geometric properties of knotted fields \cite{Bittencourt2012,Thompson:2014owa,Crisan2021,Crisan2020a} and their geometric properties \cite{Bittencourt2012, Crisan2020a, Arrayas2023}, and  new methods for constructing rational knotted fields \cite{Lechtenfeld:2017tif,Kumar:2020xjr}. The dynamics of classical test particles in knotted backgrounds was studied in \cite{Arrayas2010,Crisan2020b}, and quantum aspects of knotted fields and their wave components were presented in \cite{Crisan2020b,Arrayas2012q}. Also, the existence of field line solutions of Einstein-Maxwell equations has been investigated \cite{Vancea2017}, and reviews on knots in electromagnetism and coherent states in classical field theory have been presented \cite{Vancea2018a,Vancea2018b,Joshi2025}.

Despite these advances, the quantum description of these topological fields has remained largely unexplored.  In this paper, we present a coherent state representation for the non-null torus knot solutions. Starting from the general relation between classical initial fields and coherent state amplitudes, we construct the displacement operators and verify the defining properties through direct computation.  In the general case of Ra\~{n}ada-Hopf fields, we give the coherent states and the fundamental observables, i.e., energy, Poynting vector, helicity, photon number, quadratures, and correlation functions, as general integral formulas of complex functions determining the classical solutions. For non-null torus fields, we express the coherent states and their observables in terms of the integer parameters $(n,m,l,s)$, showing how topological information is mapped by quantization into quantum quantities.  We work throughout this paper in the Coulomb gauge with the mode expansion of the free electromagnetic field \cite{Itzykson1980,Arrayas2019}. The present work gives the first quantum description of the topological electromagnetic fields in vacuum. 

The paper is organized as follows. 
In Section~\ref{sec:coherent_states_review}, we review the canonical quantization of the electromagnetic field and the coherent states in the Coulomb gauge, and the relation between coherent state amplitudes and initial classical data.  
Also in this section, we present the explicit construction of coherent states for general Ra\~{n}ada-Hopf fields, together with the displacement operators, verification of coherent state properties, and explicit formulas for all observables.   
Section~\ref{sec:nonnull} specializes the general construction to non-null torus knots and presents closed-form expressions for the coherent states. 
In Section~\ref{sec:observables}, we give the observables of coherent states and the correlation functions. We calculate the expectation values for a non-null torus knot and particularize the results for the Hopfion defined by $n=m=l=s=1$. 
In Section~\ref{sec:conclusion}, we conclude with a discussion of our results and a presentation of future research directions.

\section{Coherent State Representation of Ra\~{n}ada-Hopf Fields}
\label{sec:coherent_states_review}

In this section, we construct the coherent states of Ra\~{n}ada-Hopf fields. After a short review of general coherent states of the electromagnetic field in vacuum, we present the Ra\~{n}ada-Hopf fields and use the general relations between classical electromagnetic fields and the corresponding coherent states to determine the coherent states and their properties.

\subsection{Coherent States of Electromagnetic Fields}

The coherent states of the electromagnetic field can be constructed from the coherent states of its normal mode oscillators \cite{Klauder1985,Itzykson1980}. Since we are interested in Ra\~{n}ada-Hopf fields, we consider the electromagnetic field in vacuum. 
In Coulomb gauge defined by $\nabla\cdot\mathbf{A}=0$ and $A_0=\Phi = 0$, the vector potential operator is expanded in normal modes as
\begin{equation}
\hat{\mathbf{A}}(\mathbf{x},t)
= 
\sum_{\lambda=1}^{2} \int \frac{d^3k}{(2\pi)^3} \frac{1}{\sqrt{2\omega_{\mathbf{k}}}} 
\left[ 
\hat{a}_{\mathbf{k}\lambda} \boldsymbol{\epsilon}_{\mathbf{k}\lambda} e^{i\mathbf{k}\cdot\mathbf{x} - i\omega_{\mathbf{k}}t} 
+ \hat{a}^{\dagger}_{\mathbf{k}\lambda} \bar{\boldsymbol{\epsilon}}_{\mathbf{k}\lambda} e^{-i\mathbf{k}\cdot\mathbf{x} + i\omega_{\mathbf{k}}t} 
\right]
\, ,
\label{eq:A_expansion}
\end{equation}
with $\omega_{\mathbf{k}} = c|\mathbf{k}|$.  
The polarization vectors $\boldsymbol{\epsilon}_{\mathbf{k}\lambda}$ are orthonormal and transverse: $\bar{\boldsymbol{\epsilon}}_{\mathbf{k}\lambda} \cdot \boldsymbol{\epsilon}_{\mathbf{k}\lambda'} = \delta_{\lambda\lambda'}$ and $\mathbf{k}\cdot\boldsymbol{\epsilon}_{\mathbf{k}\lambda}=0$.
The annihilation and creation operators satisfy the canonical commutation relations
\begin{equation}
[\hat{a}_{\mathbf{k}\lambda}, \hat{a}^{\dagger}_{\mathbf{k}'\lambda'}] = 
(2\pi)^3 \delta^{(3)}(\mathbf{k}-\mathbf{k}') \delta_{\lambda\lambda'}
\, ,
\qquad
[\hat{a}_{\mathbf{k}\lambda}, \hat{a}_{\mathbf{k}'\lambda'}] 
= [\hat{a}^{\dagger}_{\mathbf{k}\lambda}, 
\hat{a}^{\dagger}_{\mathbf{k}'\lambda'}] = 0.
\label{eq:commutation}
\end{equation}
In the Coulomb gauge, the electric and magnetic field operators follow from \eqref{eq:A_expansion} and the defining relations $\hat{\mathbf{E}} = -\partial_t\hat{\mathbf{A}}$ and $\hat{\mathbf{B}} = \nabla\times\hat{\mathbf{A}}$, and are given by 
\begin{align}
\hat{\mathbf{E}}(\mathbf{x},t) 
&= 
i\sum_{\lambda} \int \frac{d^3k}{(2\pi)^3} \sqrt{\frac{\omega_{\mathbf{k}}}{2}} 
\left[ 
\hat{a}_{\mathbf{k}\lambda} \boldsymbol{\epsilon}_{\mathbf{k}\lambda} e^{i\mathbf{k}\cdot\mathbf{x} - i\omega_{\mathbf{k}}t} 
- \hat{a}^{\dagger}_{\mathbf{k}\lambda} \bar{\boldsymbol{\epsilon}}_{\mathbf{k}\lambda} e^{-i\mathbf{k}\cdot\mathbf{x} + i\omega_{\mathbf{k}}t} 
\right]
\, , 
\label{eq:E_exp} 
\\
\hat{\mathbf{B}}(\mathbf{x},t) 
&= 
\sum_{\lambda} \int \frac{d^3k}{(2\pi)^3} \frac{1}{\sqrt{2\omega_{\mathbf{k}}}} 
\left[ 
\hat{a}_{\mathbf{k}\lambda} (i\mathbf{k}\times\boldsymbol{\epsilon}_{\mathbf{k}\lambda}) e^{i\mathbf{k}\cdot\mathbf{x} - i\omega_{\mathbf{k}}t} + \text{h.c.} 
\right]
\, . 
\label{eq:B_exp}
\end{align}
The total energy of the field is determined by the eigenvalues of the Hamiltonian operator
\begin{equation}
    \hat{H} = \frac{1}{2} \int d^3x \left( \varepsilon_0 \hat{\mathbf{E}}^2 + \frac{1}{\mu_0} \hat{\mathbf{B}}^2 \right) 
    = \sum_{\lambda} \int \frac{d^3k}{(2\pi)^3} \hbar \omega_{\mathbf{k}} \left( \hat{a}^{\dagger}_{\mathbf{k}\lambda}\hat{a}_{\mathbf{k}\lambda} + \frac{1}{2} \right).
    \label{eq:Hamiltonian}
\end{equation}

As usual, we denote the vacuum state by $|0\rangle$ satisfying
$\hat{a}_{\mathbf{k}\lambda}|0\rangle = 0$ for all $\mathbf{k}$ and $\lambda$, and the one-photon states by $|\mathbf{k},\lambda\rangle = \hat{a}^{\dagger}_{\mathbf{k}\lambda}|0\rangle$ with orthonormality relation 
\begin{equation}
\langle \mathbf{k}',\lambda'|\mathbf{k},\lambda\rangle = (2\pi)^3\delta^{(3)}(\mathbf{k}-\mathbf{k}')\delta_{\lambda\lambda'}.
\end{equation}
The single-photon states are eigenstates of the Hamiltonian with eigenvalue $\hbar\omega_{\mathbf{k}}$ above the vacuum.
Higher multi-photon states with $n$-photons and occupation numbers $n_{\mathbf{k}\lambda}$ are given by
\begin{equation}
| \{n_{\mathbf{k}\lambda}\} \rangle = \prod_{\mathbf{k},\lambda} \frac{(\hat{a}^{\dagger}_{\mathbf{k}\lambda})^{n_{\mathbf{k}\lambda}}}{\sqrt{n_{\mathbf{k}\lambda}!}}|0\rangle,
\label{eq:multi_photon_state}
\end{equation}
where the product runs over all normal modes. The multi-photon states form an orthonormal basis of the Fock space and are eigenstates of the Hamiltonian:
\begin{equation}
\hat{H}| \{n_{\mathbf{k}\lambda}\} \rangle = \sum_{\lambda}\int\frac{d^3k}{(2\pi)^3}\,\hbar\omega_{\mathbf{k}}\left(n_{\mathbf{k}\lambda}+\frac{1}{2}\right)| \{n_{\mathbf{k}\lambda}\} \rangle.
\end{equation}

Coherent states are more complex multi-particle states. For a single mode and helicity, a coherent state is defined as an eigenstate of the corresponding annihilation operator \cite{Glauber1963}
\begin{equation}
\hat{a}_{\mathbf{k}\lambda} |\{\alpha\}\rangle = \alpha_{\mathbf{k}\lambda} |\{\alpha\}\rangle 
\, , 
\label{eq:coherent_eigenvalue}
\end{equation}
for all $\mathbf{k}$ and $\lambda$, where the coherent state amplitudes $\alpha_{\mathbf{k}\lambda}$ are complex numbers. Then the coherent state of the electromagnetic field is given by 
\begin{align}
|\{\alpha\}\rangle 
&= 
\prod_{\lambda} 
\exp\left( \int \frac{d^3k}{(2\pi)^3} 
\left( \alpha_{\mathbf{k}\lambda}\hat{a}^{\dagger}_{\mathbf{k}\lambda} - \bar{\alpha}_{\mathbf{k}\lambda}\hat{a}_{\mathbf{k}\lambda} \right)\right)|0\rangle 
\nonumber
\\
&= \prod_{\lambda} 
e^{-\int \frac{d^3k}{(2\pi)^3} 
(|\alpha_{\mathbf{k}\lambda}|^2/2)} 
\prod_{\mathbf{k},\lambda} 
\left( \sum_{n_{\mathbf{k}\lambda}=0}^{\infty} \frac{\alpha_{\mathbf{k}\lambda}^{n_{\mathbf{k}\lambda}}}{\sqrt{n_{\mathbf{k}\lambda}!}} |n_{\mathbf{k}\lambda}\rangle
\right).
\label{eq:coherent_fock_expansion}
\end{align}
The expansion on the right-hand side of equation \eqref{eq:coherent_fock_expansion} shows that coherent states have Poissonian photon number statistics. These states can be generated from the vacuum by the displacement operator
\begin{equation}
    \hat{D}(\{\alpha\}) = \exp\left[ \sum_{\lambda} \int \frac{d^3k}{(2\pi)^3} \left( \alpha_{\mathbf{k}\lambda} \hat{a}^{\dagger}_{\mathbf{k}\lambda} - \bar{\alpha}_{\mathbf{k}\lambda} \hat{a}_{\mathbf{k}\lambda} \right) \right],
    \label{eq:displacement}
\end{equation}
so that $|\{\alpha\}\rangle = \hat{D}(\{\alpha\})|0\rangle$. The displacement operator is unitary and satisfies
\begin{equation}
    \hat{D}^{\dagger}(\{\alpha\}) \hat{a}_{\mathbf{k}\lambda} \hat{D}(\{\alpha\}) = \hat{a}_{\mathbf{k}\lambda} + \alpha_{\mathbf{k}\lambda}.
    \label{eq:displacement_action}
\end{equation}
Coherent states saturate the Heisenberg uncertainty relation for each mode; under free evolution they remain coherent, with their amplitude acquiring a time-proportional phase; and the expectation values of field operators in a coherent state equal the classical fields constructed from the amplitudes $\alpha_{\mathbf{k}\lambda}$.

For a given classical solution $\mathbf{E}_{\mathrm{cl}}(\mathbf{x},t)$, $\mathbf{B}_{\mathrm{cl}}(\mathbf{x},t)$, the corresponding coherent state amplitudes are determined by the initial data at $t=0$ \cite{Glauber1963}. One can easily show that
\begin{equation}
\alpha_{\mathbf{k}\lambda} = \frac{1}{\sqrt{2\omega_{\mathbf{k}}}}\,
\bar{\boldsymbol{\epsilon}}_{\mathbf{k}\lambda} \cdot \int d^3x\, e^{-i\mathbf{k}\cdot\mathbf{x}} \Bigl( \omega_{\mathbf{k}} \mathbf{A}_{\mathrm{cl}}(\mathbf{x},0) - i \mathbf{E}_{\mathrm{cl}}(\mathbf{x},0) \Bigr)
\, .
\label{eq:alpha_from_fields}
\end{equation}
Here, $\mathbf{A}_{\mathrm{cl}}$ is any vector potential in Coulomb gauge satisfying $\mathbf{B}_{\mathrm{cl}} = \nabla\times\mathbf{A}_{\mathrm{cl}}$.  This formula ensures that the expectation value of $\hat{\mathbf{E}}$ in the coherent state $|\{\alpha\}\rangle$ reproduces $\mathbf{E}_{\mathrm{cl}}$ at all times. 

\subsection{Coherent States of Ra\~{n}ada-Hopf Fields}

Consider two complex scalar maps $\phi, \theta : \mathbb{R}^3\cup\{\infty\}\cong S^3 \to S^2$ representing Hopf fibrations at any fixed time $t$. The level sets $S^1$ of $\phi$, respectively $\theta$, are linked with each other. The Ra\~{n}ada–Trueba electric and magnetic fields are given by \cite{Ranada1992,Ranada1995, Arrayas2015}
\begin{align}
\mathbf{E}_{\mathrm{cl}}(\mathbf{x},t) &= \frac{\sqrt{a}\,c}{2\pi i}\,\frac{\nabla\theta (\mathbf{x},t)\times \nabla\bar{\theta} (\mathbf{x},t)}{(1+|\theta (\mathbf{x},t)|^2 )^2}, \label{eq:E_ranada}
\\
\mathbf{B}_{\mathrm{cl}}(\mathbf{x},t) &= \frac{\sqrt{a}}{2\pi i}\,\frac{\nabla\phi (\mathbf{x},t) \times \nabla\bar{\phi} (\mathbf{x},t)}{(1+|\phi (\mathbf{x},t)|^2)^2}, \label{eq:B_ranada}
\end{align}
where $\sqrt{a}$ is a constant with dimensions that ensure the correct SI units. The fields $\mathbf{E}_{\mathrm{cl}}$ and $\mathbf{B}_{\mathrm{cl}}$ satisfy Maxwell's equations in vacuum and have localized energy density. 
The Ra\~{n}ada-Hopf solution from equations \eqref{eq:E_ranada} and \eqref{eq:B_ranada} is the prototypical example of a classical topological vacuum electromagnetic field. 

In order to construct the quantum coherent state representation of the Ra\~{n}ada-Hopf field, we can employ the relations \eqref{eq:displacement} and \eqref{eq:alpha_from_fields} and obtain 
\begin{align}
|\{\alpha\}\rangle_{RH} 
&= 
\hat{D}_{\mathrm{RH}} | 0 \rangle 
\nonumber
\\
&= \exp
\Bigg\{ 
	\frac{\sqrt{a}}{(2\pi)^4} 
	\int \frac{d^3k}{\sqrt{2\omega_k}}
	\sum_{\lambda} 
		\Bigg[ 
			\bar{\boldsymbol{\epsilon}}_{\mathbf{k}\lambda}
			\cdot \displaystyle\int d^3x \, 
			e^{-i\mathbf{k}\cdot\mathbf{x}} \omega_k
			\left( 
				\frac{\phi\nabla\bar\phi 
				- 
				\bar\phi\nabla\phi}{(1+|\phi|^2)^2} 
				- 
				\frac{c}{\omega_k} \frac{\nabla\theta 
				\times \nabla\bar\theta}{(1+|\theta|^2)^2} 
			\right)_{\!t=0}
				\hat{a}^{\dagger}_{\mathbf{k}\lambda} 
%\right.
%\right.				
\nonumber
\\
&- 
%\left.
%\left.
\mathrm{h.c.} 
\Bigg] 
\Bigg\} | 0 \rangle
				\, .
    \label{eq:D_hopf_explicit}
\end{align}
As before, the dot denotes the three-dimensional scalar product. Equation \eqref{eq:D_hopf_explicit} is the most general integral representation of the coherent Ra\~{n}ada-Hopf state in of terms general Hopf maps and normal modes. We can see from \eqref{eq:D_hopf_explicit} that the Ra\~{n}ada-Hopf coherent state amplitude has the following form
\begin{equation}
\alpha_{\mathbf{k}\lambda} = 
\frac{\sqrt{a}}{(2\pi)^4\sqrt{2\omega_k}} \bar{\boldsymbol{\epsilon}}_{\mathbf{k}\lambda} \cdot 
        \int d^3x \, e^{-i\mathbf{k}\cdot\mathbf{x}} 
		\omega_k        
        \left( 
	        \frac{\phi\nabla\bar{\phi} 
    	    - \bar{\phi}\nabla\phi}{(1+|\phi|^2)^2} 
        	- \frac{c}{\omega_k} \frac{\nabla\theta
        	 \times
        	\nabla\bar{\theta}}{(1+|\theta|^2)^2} 
         \right)_{t=0}.
\label{eq:alpha_hopfion}
\end{equation}
In equation \eqref{eq:alpha_hopfion}, the integrals are well-defined due to the rapid decay of the fields as 
$|\mathbf{x}|\to\infty$.  

We can define the observables for $|\{\alpha\}\rangle_{RH}$ and compute their expectation values in the standard way. 
The field expectation values in $|\{\alpha\}\rangle_{RH}$ can be derived using the properties of the helicity basis and of the creation and annihilation operators, and we obtain
\begin{align}
\mathbf{E}_{\mathrm{cl}}(\mathbf{x},t) 
& =
\langle \hat{\mathbf{E}}(\mathbf{x},t) \rangle 
= \frac{\sqrt{a}\,c}{2\pi i}\,\frac{\nabla\theta \times \nabla\bar{\theta}}{(1+|\theta|^2)^2},
\label{eq:hopfion_E_expval}
\\
\mathbf{B}_{\mathrm{cl}}(\mathbf{x},t) 
& =
\langle \hat{\mathbf{B}}(\mathbf{x},t) \rangle 
= \frac{\sqrt{a}}{2\pi i}\,\frac{\nabla\phi \times \nabla\bar{\phi}}{(1+|\phi|^2)^2} .
\label{eq:hopfion_B_expval}
\end{align}
Equations \eqref{eq:hopfion_E_expval} and \eqref{eq:hopfion_B_expval} show that the expectation values of the quantum electromagnetic field in the coherent state $|\{\alpha\}\rangle_{RH}$ are exactly the classical Ra\~{n}ada-Hopf fields, which is a consistency check of our construction.
%Photon number and energy
From the definition of the number operator and Hamiltonian of the electromagnetic field, we obtain the photon number and total energy densities as usual. The expectation value of the photon number operator in mode $(\mathbf{k},\lambda)$ is given by
\begin{equation}
\langle \hat{N}_{\mathbf{k}\lambda} \rangle = \langle \hat{a}^{\dagger}_{\mathbf{k}\lambda} \hat{a}_{\mathbf{k}\lambda} \rangle = |\alpha_{\mathbf{k}\lambda}|^2
\, ,
\label{eq:photon_number}
\end{equation}
which allows us to calculate the total photon number with normal ordering
\begin{equation}
\langle :\!\hat{N}\!:\rangle = \sum_{\lambda} \int \frac{d^3k}{(2\pi)^3} |\alpha_{\mathbf{k}\lambda}|^2
\, ,
\label{eq:total_photon_number}
\end{equation}
and the energy expectation value in the coherent state
\begin{equation}
\langle :\!\hat{H}\!:\rangle = \sum_{\lambda} \int \frac{d^3k}{(2\pi)^3} \hbar \omega_{\mathbf{k}} |\alpha_{\mathbf{k}\lambda}|^2
\, .
\label{eq:energy_expectation}
\end{equation}
In these relations, $\alpha_{\mathbf{k}\lambda}$ is given by equation \eqref{eq:alpha_hopfion}. Similarly, using the normally ordered expressions for energy and Poynting vector densities
\begin{align}
:\!\hat{u}(\mathbf{x},t)\!: 
& = 
\frac{\varepsilon_0}{2} :\!\hat{\mathbf{E}}^2(\mathbf{x},t)\!: + \frac{1}{2\mu_0} :\!\hat{\mathbf{B}}^2(\mathbf{x},t)\!: \, ,
\label{eq:energy_density_op}
\\
:\!\hat{\mathbf{S}}(\mathbf{x},t)\!: 
& = 
\frac{1}{\mu_0} :\!\hat{\mathbf{E}}(\mathbf{x},t) \times \hat{\mathbf{B}}(\mathbf{x},t)\!: \, ,
\label{eq:poynting_op}
\end{align}
we get
\begin{align}
u_{\mathrm{cl}}(\mathbf{x},t) 
& =
\langle : \!\hat{u}(\mathbf{x},t)\!:  \rangle
= 
\frac{a}{8\mu_0\pi^2} 
\left[ \frac{|\nabla\phi\times\nabla\bar{\phi}|^2}{(1+|\phi|^2)^4} 
+ 
\frac{|\nabla\theta\times\nabla\bar{\theta}|^2}{(1+|\theta|^2)^4} \right]
\, ,
\label{eq:hopfion_energy_density}
\\
{\mathbf S}_{\mathrm{cl}}(\mathbf{x},t) 
& =
\langle :\!\hat{\mathbf{S}}(\mathbf{x},t)\!:\rangle 
= 
- \frac{a c}{4\pi^2\mu_0 }  \frac{(\nabla\theta\times\nabla\bar{\theta})\times(\nabla\phi\times\nabla\bar{\phi})}{(1+|\theta|^2)^2(1+|\phi|^2)^2}
\, .
\label{eq:hopfion_poynting}
\end{align}
We define the quadrature operators as usual by
\begin{equation}
\hat{X}_{\mathbf{k}\lambda} 
= \frac{1}{\sqrt{2}}(\hat{a}_{\mathbf{k}\lambda} + \hat{a}^{\dagger}_{\mathbf{k}\lambda}),
\quad
\hat{P}_{\mathbf{k}\lambda} 
= 
\frac{1}{i\sqrt{2}}(\hat{a}_{\mathbf{k}\lambda} - \hat{a}^{\dagger}_{\mathbf{k}\lambda}),
\label{eq:quadratures}
\end{equation}
for every mode $(\mathbf{k},\lambda)$.
These operators satisfy the canonical commutation relations $[\hat{X},\hat{P}] = i$ (in natural units $\hbar =1$). Then $|\{\alpha\}\rangle_{RH}$ minimizes the Heisenberg uncertainty relation
\begin{equation}
(\Delta X_{\mathbf{k}\lambda})^2 = (\Delta P_{\mathbf{k}\lambda})^2 = \frac{1}{2},
\qquad 
(\Delta X_{\mathbf{k}\lambda}) (\Delta P_{\mathbf{k}\lambda}) 
= \frac{1}{2}.
\label{eq:hopfion_uncertainties}
\end{equation}

The first-order and second-order correlation functions are obtained from the positive-frequency part of the electric field, $\mathbf{E}_{\mathrm{cl}}^{(+)}$. This is the analytic component corresponding to $\mathbf{E}_{\mathrm{cl}}$ and can be read off from equation \eqref{eq:E_exp}. One can show that in the coherent state $|\{\alpha\}\rangle_{RH}$, the correlation functions are \cite{Glauber1963}
\begin{align}
G^{(1)}(x,y) &= \bigl(\mathbf{E}_{\mathrm{cl}}^{(+)}(x)\bigr)^* \!\cdot\! \mathbf{E}_{\mathrm{cl}}^{(+)}(y),
\label{eq:corr_fct_1}
\\
G^{(2)}(x,y) &= |\mathbf{E}_{\mathrm{cl}}^{(+)}(x)|^2 |\mathbf{E}_{\mathrm{cl}}^{(+)}(y)|^2,
\label{eq:corr_fct_2}
\end{align}
from which we get the normalized correlations
\begin{align}
g^{(1)}(x,y) 
& = 
\frac{G^{(1)}(x,y)}{\sqrt{G^{(1)}(x,x)G^{(1)}(y,y)}} 
\, ,
\label{eq:hopfion_g1_final}
\\
g^{(2)}(x,y) & = 
\frac{G^{(2)}(x,y)}{\langle \hat{I}(x)\rangle \langle \hat{I}(y)\rangle} 
\, ,
\label{eq:hopfion_g2_final}
\end{align}
where $\hat{I}(x) = \hat{\mathbf{E}}^{(-)}(x)\cdot\hat{\mathbf{E}}^{(+)}(x)$.
In general, the functions from equations \eqref{eq:hopfion_g1_final} and \eqref{eq:hopfion_g2_final} should be determined for each class of topological coherent states.

\subsection{Time evolution of coherent states}

The time evolution of the coherent state is described by the free Hamiltonian of the electromagnetic field
\begin{equation}
|\{\alpha\}(t)\rangle = e^{-i\hat{H}t/\hbar} |\{\alpha\}\rangle
\, .
\label{eq:time_ev_state}
\end{equation}
Because $\hat{H}$ is quadratic in creation and annihilation operators, it is well known that a coherent state remains coherent under time evolution \cite{Itzykson1980}, with the amplitudes acquiring a phase proportional to frequency
\begin{equation}
e^{-i\hat{H}t/\hbar} \hat{a}_{\mathbf{k}\lambda} e^{i\hat{H}t/\hbar} = \hat{a}_{\mathbf{k}\lambda} e^{-i\omega_{\mathbf{k}}t}
\label{eq:a_time_phase}
\, .
\end{equation}
Therefore, the expectation values of the field operators at time $t$ are given by
\begin{align}
\langle \hat{\mathbf{E}}(\mathbf{x},t)\rangle &= i\sum_{\lambda} \int \frac{d^3k}{(2\pi)^3} \sqrt{\frac{\omega_{\mathbf{k}}}{2}} 
\left[ \alpha_{\mathbf{k}\lambda} e^{-i\omega_{\mathbf{k}}t} \boldsymbol{\epsilon}_{\mathbf{k}\lambda} e^{i\mathbf{k}\cdot\mathbf{x}} - \text{c.c.} \right] = \mathbf{E}_{\mathrm{cl}}(\mathbf{x},t)
\, ,
\label{eq:E_time}
\\
\langle \hat{\mathbf{B}}(\mathbf{x},t)\rangle &= \sum_{\lambda} \int \frac{d^3k}{(2\pi)^3} \frac{1}{\sqrt{2\omega_{\mathbf{k}}}} 
\left[ \alpha_{\mathbf{k}\lambda} e^{-i\omega_{\mathbf{k}}t} (i\mathbf{k}\times\boldsymbol{\epsilon}_{\mathbf{k}\lambda}) e^{i\mathbf{k}\cdot\mathbf{x}} + \text{c.c.} \right] = \mathbf{B}_{\mathrm{cl}}(\mathbf{x},t)
\label{eq:B_time}
\, .
\end{align}
where $\mathbf{E}_{\mathrm{cl}}(\mathbf{x},t)$ and $\mathbf{B}_{\mathrm{cl}}(\mathbf{x},t)$ are the classical fields evolved from the initial data. 

Since a coherent state is completely determined by its eigenvalues $\alpha_{\mathbf{k}\lambda}$, we have 
\begin{equation}
\hat{a}_{\mathbf{k}\lambda}|\{\alpha\}(t)\rangle = \alpha_{\mathbf{k}\lambda}e^{-i\omega_{\mathbf{k}}t}|\{\alpha\}(t)\rangle
\, .
\label{eq:a_at_t}
\end{equation}

For a classical electromagnetic field $\mathbf{E}_{\mathrm{cl}}(\mathbf{x},t),\mathbf{B}_{\mathrm{cl}}(\mathbf{x},t)$ that satisfies Maxwell’s equations in vacuum, the standard construction of a coherent state that reproduces this field at a given time 
$t_0$ uses the initial data at that time
\begin{equation}
\alpha_{\mathbf{k}\lambda}(t_0) = \frac{1}{\sqrt{2\omega_{\mathbf{k}}}}\,
\bar{\boldsymbol{\epsilon}}_{\mathbf{k}\lambda}\!\cdot\!\int d^3x\,e^{-i\mathbf{k}\cdot\mathbf{x}}\Bigl(\omega_{\mathbf{k}}\mathbf{A}_{\mathrm{cl}}(\mathbf{x},t_0)-i\mathbf{E}_{\mathrm{cl}}(\mathbf{x},t_0)\Bigr)
\, .
\label{eq:a_t_init}
\end{equation}
If we take $t_0=0$, we obtain the amplitudes $\alpha_{\mathbf{k}\lambda}(0)=\alpha_{\mathbf{k}\lambda}$.  
Then the time-evolved state has mode amplitudes $\alpha_{\mathbf{k}\lambda}(t)=\alpha_{\mathbf{k}\lambda}e^{-i\omega_{\mathbf{k}}t}$.

On the other hand, if we directly apply the same formula at a later time \(t\), using the classical fields at that time, we obtain
\begin{equation}
\alpha_{\mathbf{k}\lambda}^{\mathrm{(direct)}}(t) = \frac{1}{\sqrt{2\omega_{\mathbf{k}}}}\,
\bar{\boldsymbol{\epsilon}}_{\mathbf{k}\lambda}\!\cdot\!\int d^3x\,e^{-i\mathbf{k}\cdot\mathbf{x}}\Bigl(\omega_{\mathbf{k}}\mathbf{A}_{\mathrm{cl}}(\mathbf{x},t)-i\mathbf{E}_{\mathrm{cl}}(\mathbf{x},t)\Bigr)
\, .
\label{eq:a_time_direct}
\end{equation}
It is straightforward to show that
\begin{equation}
\alpha_{\mathbf{k}\lambda}^{\mathrm{(direct)}}(t) = \alpha_{\mathbf{k}\lambda}(0)\,e^{-i\omega_{\mathbf{k}}t}
\, .
\label{eq:equiv_direct}
\end{equation}
Hence the two constructions give exactly the same time-dependent amplitudes.  Consequently, the coherent state built directly from the classical fields at time \(t\) is identical to the time-evolved state obtained from the initial coherent state.

\section{Coherent States for Non-Null Torus Knots}
\label{sec:nonnull}

In this section, we apply the construction of coherent states to the general class of non-null torus knots parameterized by two pairs of coprime integers $(n,m)$ and $(l,s)$. This family contains as a particular case the null torus knot solutions when all parameters are equal to each other \cite{Arrayas2015, Arrayas2017}. 

Consider a family of non-null electromagnetic fields parameterized by four positive integers $(n,m,l,s)$ \cite{Arrayas2015, Arrayas2017}. At initial time, the magnetic and electric fields are given explicitly by
\begin{align}
\mathbf{E}(\mathbf{x},0) 
& = 
\frac{8c\sqrt{a}}{\pi L_0^2 (1+R^2)^3} \begin{pmatrix} l\displaystyle\frac{X^2-Y^2-Z^2+1}{2} \\ lXY - sZ \\ lXZ + sY \end{pmatrix}
\, ,
\label{eq:E0_nonnull}
\\
\mathbf{B}(\mathbf{x},0) 
& = 
\frac{8\sqrt{a}}{\pi L_0^2 (1+R^2)^3} \begin{pmatrix} mY - nXZ \\ -mX - nYZ \\ n\displaystyle\frac{X^2+Y^2-Z^2-1}{2} \end{pmatrix}
,
\label{eq:B0_nonnull}
\end{align}
Here, we use use for convenience the dimensionless variables introduced in \cite{Arrayas2015,Arrayas2017}: $(X,Y,Z) = (x,y,z)/L_0$, $R^2 = X^2+Y^2+Z^2$, $ct/L_0 = T$, with $L_0$ a constant with dimensions of length.
The fields $\mathbf{B}(\mathbf{x},0)$ and $\mathbf{E}(\mathbf{x},0)$ are exact solutions of Maxwell's equations in vacuum. The magnetic lines at $T=t=0$ are $(n,m)$ torus knots with linking number $nm$, while the electric lines are $(l,s)$ torus knots with linking number $ls$.

The full time-dependent fields can be obtained by applying the Fourier transform \cite{Arrayas2019}. We define the complex combination
\begin{equation}
\mathbf{F}(\mathbf{k}) = \frac{1}{(2\pi)^{3/2}} \int d^3 x \, 
e^{i\mathbf{k}\cdot\mathbf{x}} 
\left( \mathbf{B}(\mathbf{x},0) + \frac{i}{c}\mathbf{E}(\mathbf{x},0) 
\right)
\, ,
\label{eq:F_nonnull}
\end{equation}
and the time-dependent fields are
\begin{equation}
\mathbf{B}(\mathbf{x},t) + \frac{i}{c}\mathbf{E}(\mathbf{x},t) 
= \frac{1}{(2\pi)^{3/2}} \int d^3 k \, e^{-i\mathbf{k}\cdot\mathbf{x}} 
\mathbf{F}(\mathbf{k}) e^{-i\omega_{\mathbf{k}}t}
\, .
\label{eq:complex_field_time}
\end{equation}
In \cite{Arrayas2015}, the electric and magnetic fields were shown to have the following compact form
\begin{align}
\mathbf{E}(\mathbf{x},t) 
& = 
\frac{\sqrt{a}c}{\pi L_0^2} \, \frac{Q\mathbf{H}_4 - P\mathbf{H}_3}{(A^2+T^2)^3} 
\, ,
\label{eq:E_nonnull_time}
\\
\mathbf{B}(\mathbf{x},t) 
& = 
\frac{\sqrt{a}}{\pi L_0^2} 
\, 
\frac{Q\mathbf{H}_1 + P\mathbf{H}_2}{(A^2+T^2)^3} 
\, ,
\label{eq:B_nonnull_time}
\end{align}
where we use the shorthand notation
\begin{align}
A &= \frac{R^2 - T^2 + 1}{2}
\, ,
\label{eq:A_def} 
\\
P &= T(T^2 - 3A^2)
\, ,
\label{eq:P_def} 
\\
Q &= A(A^2 - 3T^2)
\, . 
\label{eq:Q_def}
\end{align}
The vectors $\mathbf{H}_i$ are given by
\begin{align}
\mathbf{H}_1 
&= \begin{pmatrix}
-nXZ + mY + sT 
\\
-nYZ - mX - lTZ 
\\
n\frac{X^2+Y^2-Z^2-1+T^2}{2} + lTY
\end{pmatrix}
, 
\label{eq:H1} 
\\
\mathbf{H}_2 &= \begin{pmatrix}
s\frac{1+X^2-Y^2-Z^2-T^2}{2} - mTY 
\\
sXY - lZ + mTX 
\\
sXZ + lY + nT
\end{pmatrix}
, 
\label{eq:H2} 
\\
\mathbf{H}_3 &= \begin{pmatrix}
-mXZ + nY + lT 
\\
-mYZ - nX - sTZ 
\\
m\frac{X^2+Y^2-Z^2-1+T^2}{2} + sTY
\end{pmatrix}
, 
\label{eq:H3} 
\\
\mathbf{H}_4 &= \begin{pmatrix}
l\frac{1+X^2-Y^2-Z^2-T^2}{2} - nTY 
\\
lXY - sZ + nTX 
\\
lXZ + sY + mT
\end{pmatrix}
. 
\label{eq:H4}
\end{align}

To obtain the coherent state amplitudes $\alpha_{\mathbf{k}\lambda}$ corresponding to the classical field configuration from equations \eqref{eq:E_nonnull_time} and \eqref{eq:B_nonnull_time}, we need the vector potential $\mathbf{A}_{\mathrm{cl}}(\mathbf{x},0)$. In the Coulomb gauge and for the non-null torus knot family given above, we can compute the vector potential through its Fourier transform as \cite{Arrayas2015}
\begin{equation}
\mathbf{A}_{\mathrm{cl}}(\mathbf{x},0) = \frac{1}{(2\pi)^{3/2}} \int d^3k \, e^{i\mathbf{k}\cdot\mathbf{x}} \tilde{\mathbf{A}}(\mathbf{k})
\, ,
\label{eq:A_fourier}
\end{equation}
where $\tilde{\mathbf{A}}(\mathbf{k})$ is related to the Fourier transform of the initial fields. After some lengthy but straightforward calculations, we obtain the coefficient amplitudes defined in equation \eqref{eq:alpha_from_fields} above
\begin{equation}
\alpha_{\mathbf{k}\lambda} 
= 
\frac{\sqrt{a} L_0}{2^{3/2}\pi^2 \sqrt{2\omega_{\mathbf{k}}}} \, e^{-K} \,
\boldsymbol{\epsilon}_{\mathbf{k}\lambda}^* \cdot \mathbf{W}(\mathbf{K}) 
\, .
\label{eq:alpha_closed}
\end{equation}
Here, the vector $\mathbf{W}(\mathbf{K})$ is given by
\begin{equation}
\mathbf{W}(\mathbf{K}) = \begin{pmatrix}
        i m K_y + n \dfrac{K_x K_z}{K} + i l 
        \left(
        K - \dfrac{K_x^2}{K}
        \right) 
        \\
        -i m K_x + n \dfrac{K_y K_z}{K} - i l \dfrac{K_x K_y}{K} + s K_z 
        \\
        n\left(
        -K + \dfrac{K_z^2}{K}
        \right) - i l \dfrac{K_x K_z}{K} - s K_y
    \end{pmatrix}
\, ,    
\label{eq:W_def}
\end{equation}
and we use the conjugate variables $\mathbf{K} = L_0 \mathbf{k}$ with $K=|\mathbf{K}|$. The exponential factor $e^{-K}$ in equation \eqref{eq:alpha_closed} shows that the mode amplitudes are strongly suppressed for wavelengths smaller than $L_0$ ($k > 1/L_0$), which is a consequence of the finite size of the topological structure. From equation \eqref{eq:alpha_closed}, we construct the displacement operator that takes the following form
\begin{equation}
\hat{D}_{\mathrm{knot}} 
= 
\exp
\left[ 
\frac{\sqrt{a}}{2^{3/2}\pi^2 L_0^2} \sum_{\lambda} \int \frac{d^3 K}{(2\pi)^3} \frac{e^{-K}}{\sqrt{2\omega_{\mathbf{K}}}} 
\times 
\left( 
	\left(
	\boldsymbol{\epsilon}_{\mathbf{k}\lambda}^* \cdot \mathbf{W}(\mathbf{K})
	\right) 
	\hat{a}^{\dagger}_{\mathbf{k}\lambda} 
	- 
	\left(
	\boldsymbol{\epsilon}_{\mathbf{k}\lambda} \cdot \mathbf{W}^*(\mathbf{K})
	\right) \hat{a}_{\mathbf{k}\lambda} 
\right) 
\right]
\, ,
\label{eq:D_final_closed}
\end{equation}
with $\omega_{\mathbf{K}} = L_0 \omega_{\mathbf{k}}$. Equation \eqref{eq:D_final_closed} contains only elementary functions of the wave vector $\mathbf{K}$ and the mode operators.  The exponential factor $e^{-K}$ ensures that the integral converges at high energies and is a consequence of the finite spatial extent of the classical knotted field. The vector $\mathbf{W}$ encodes the topological information about the knot through the integers $(n,m,l,s)$, as well as the parameters $a$ and $L_0$ of the classical field.

The displacement operator generates the coherent state by acting on the electromagnetic vacuum as usual
\begin{equation}
|\mathrm{knot}\rangle = \hat{D}_{\mathrm{knot}}|0\rangle
\, .
\label{eq:coherent_non_null_knot}
\end{equation} 
All observable eigenvalues in the knotted coherent states can be computed from it using the standard properties of coherent states and the explicit amplitudes from equation \eqref{eq:alpha_closed} given above. We will focus on these observables in the next section.

\section{Field Observables in Coherent States of Non-Null Torus Knots}
\label{sec:observables}

The coherent states $|\mathrm{knot}\rangle$ given by equation \eqref{eq:coherent_non_null_knot} can be used to determine the field observables. We compute explicitly the energy, Poynting vector, photon helicity, and photon number in these states.

% Energy 
The Hamiltonian operator for the electromagnetic field has the standard form in terms of annihilation and creation operators
\begin{equation}
\hat{H} = 
\sum_\lambda \int \frac{d^3k}{(2\pi)^3} \hbar\omega_{\mathbf{k}} 
\left( 
\hat{a}^\dagger_{\mathbf{k}\lambda}\hat{a}_{\mathbf{k}\lambda} + \frac12 
\right)
\, .
\label{eq:Hamiltonian_sec4}
\end{equation}
Using equation \eqref{eq:alpha_closed}, we can show that its expectation value in the coherent state $|\mathrm{knot}\rangle$, after normal ordering, is given by 
\begin{equation}
\langle :\! \hat{H} \! : \rangle
= \frac{a}{8\pi^4 L_0^2} \int \frac{d^3 K}{(2\pi)^3} \, K \, e^{-2K} \sum_{\lambda} 
\bigl| 
\boldsymbol{\epsilon}_{\mathbf{k}\lambda}^* \cdot \mathbf{W}(\mathbf{K}) 
\bigr|^2
\, .
\label{eq:H_expect}
\end{equation}
The polarization sum and the integral can be evaluated to obtain
\begin{equation}
\langle :\!\hat{H}\!: \rangle 
= \frac{\hbar c a}{32\pi^6 }\,(n^2+m^2+l^2+s^2)
\, .
\label{eq:H_expect_1}
\end{equation}
This shows that the energy of the knotted coherent state is determined by the combination $n^2+m^2+l^2+s^2$, which means that the topological toroidal configurations are degenerate in energy.

%Linear momentum
Let us compute the linear momentum density. The normally ordered Poynting vector operator is defined by
\begin{equation}
:\!\hat{\mathbf{S}}\!: = \frac{1}{\mu_0} :\!\hat{\mathbf{E}}\times\hat{\mathbf{B}}\!:
\, ,
\label{eq:S_expect}
\end{equation}
and we can show that its expectation in the coherent state equals the classical Poynting vector
\begin{equation}
\langle :\!\hat{\mathbf{S}}(\mathbf{X},T)\!:\rangle = \frac{1}{\mu_0} \mathbf{E}_{\mathrm{cl}}(\mathbf{X},T) \times \mathbf{B}_{\mathrm{cl}}(\mathbf{X},T)
\, .
\label{eq:S_classical}
\end{equation}
Using the explicit fields given in equations \eqref{eq:E_nonnull_time} and \eqref{eq:B_nonnull_time}, we obtain after some lengthy computations
\begin{equation}
\langle :\!\hat{\mathbf{S}} (\mathbf{X},T) \!:\rangle 
= \frac{2 a c (n^2+m^2+l^2+s^2)}{\mu_0\pi^2 L_0^4 (1+R^2+T^2)^4}\,
\begin{pmatrix}
TX \\ TY \\ TZ - \frac{1+R^2-T^2}{2}
\end{pmatrix}
\, .
\label{eq:S_expect_1}
\end{equation}
Equation \eqref{eq:S_expect_1} shows that the expectation value of the Poynting vector depends on the integers only through the same factor $(n^2+m^2+l^2+s^2)$ and exhibits a toroidal structure and a momentum density topological degeneracy. At initial time $T=0$, the vector $\langle :\!\hat{\mathbf{S}} (\mathbf{X},0) \!:\rangle$ is oriented along the $Z$-axis in the reference frame defined by the torus, acquiring components along the other axes in time.

%Helicity
Now let us discuss the helicity, another important observable of the quantum electromagnetic field. In quantum theory, there are two distinct helicity operators: $\hat{\Lambda}$, which is the optical or spin helicity, and $\hat{H}_{\mathrm{m}}$, the magnetic helicity. We discuss both in what follows.  

\paragraph{Photon helicity} 
For a free electromagnetic field in the Coulomb gauge, the (optical) helicity operator is defined as
\begin{equation}
\hat{\Lambda} = \int d^3x \, \hat{\mathbf{E}}(\mathbf{x})\cdot\hat{\mathbf{B}}(\mathbf{x})
\label{eq:Spin_optical}
\, ,
\end{equation}
where the electric and magnetic fields are expressed in terms of the vector potential 
$\hat{\mathbf{A}}$ as $\hat{\mathbf{E}} = -\partial_t\hat{\mathbf{A}}$ and $\hat{\mathbf{B}} = \nabla\times\hat{\mathbf{A}}$. The photon helicity measures the circular polarization content of the field and is proportional to the difference in photon number between right- and left-handed circularly polarized modes. The operator $\hat{\Lambda}$ is gauge invariant for free fields. For monochromatic fields, $\hat{\Lambda}$ coincides, up to the factor $\omega_{\mathbf{k}}$, with the operator $\int \hat{\mathbf{A}}(\mathbf{x} ) \cdot\hat{\mathbf{B}} (\mathbf{x} ) \,d^3x$. One can show that in the circular polarization basis, the operator $\hat{\Lambda}$ has the following form
\begin{equation}
: \hat{\Lambda} : \, = \sum_{\lambda=\pm} \lambda \int \frac{d^3k}{(2\pi)^3} \hat{a}^\dagger_{\mathbf{k}\lambda}\hat{a}_{\mathbf{k}\lambda}
\, .
\label{eq:Spin_operator}
\end{equation}
We can compute the expectation value in the knot coherent state exactly, arriving at
\begin{equation}
\langle \, : \hat{\Lambda} : \, \rangle = \frac{a}{32\pi^6 L_0^2}\,(mn+ls)
\, .
\label{eq:Spin_expect}
\end{equation}
The right-hand side of equation \eqref{eq:Spin_expect} has an overall sign that depends on the convention for the circular polarization basis. To obtain the above formula, we used the standard definition $\boldsymbol{\epsilon}_{\mathbf{k}+}=(\hat{\mathbf{e}}_1+i\hat{\mathbf{e}}_2)/\sqrt{2}$.

Equation \eqref{eq:Spin_expect} shows that the expectation value of photon helicity is determined by $mn$ and $ls$, as expected from classical computations \cite{Arrayas2017}.

\paragraph{Magnetic helicity} The magnetic helicity operator is defined as 
\begin{equation}
\hat{H}_{\mathrm{m}} = 
\int d^3x \,\hat{\mathbf{A}} (\mathbf{x} ) \!
\cdot\!\hat{\mathbf{B}} (\mathbf{x})
\, .
\label{eq:Helicity_magnetic_def}
\end{equation}
The operator $\hat{H}_{\mathrm{m}}$ is a measure of the linking and twisting of magnetic field lines and is a topological invariant in ideal magnetohydrodynamics. The expectation value of $\hat{H}_{\mathrm{m}}$ can be computed similarly to that of the spin operator. After some calculations we obtain, up to a conventional sign which depends on the choice of circular polarization basis,
\begin{equation}
\langle \, :  \hat{H}_{\mathrm{m}} : \, \rangle = \frac{a}{160\pi^6 L_0}\,(mn+ls)
\, .
\label{eq:Helicity_magnetic_comp}
\end{equation}
Comparing equations \eqref{eq:Spin_expect} and \eqref{eq:Helicity_magnetic_comp}, we see that the same combination of integers determines the expectation values of both helicity operators in the knotted coherent state. The distinction between the two lies in their physical dimensions and different scaling behaviours with the typical length scale $L_0$.

%Photon number
It is interesting to determine the photon number in the knotted coherent state. The normally ordered total photon number operator is defined by 
\begin{equation}
: \hat{N} : = \sum_\lambda \int \frac{d^3k}{(2\pi)^3} \hat{a}^\dagger_{\mathbf{k}\lambda}\hat{a}_{\mathbf{k}\lambda}
\, ,
\label{eq:N_def}
\end{equation}
with expectation value given by
\begin{equation}
\langle \, : \hat{N} : \,  \rangle = \sum_\lambda \int \frac{d^3k}{(2\pi)^3} |\alpha_{\mathbf{k}\lambda}|^2
\, .
\label{eq:N_exp_val}
\end{equation}
Using the same polarization sum as for the energy, we obtain 
\begin{equation}
\langle \, : \hat{N} \, : \rangle = 
\frac{a}{8\pi^4 cL_0^2} \int \frac{d^3 K}{(2\pi)^3} \, K \, e^{-2K} \left( |\mathbf{W}|^2 - \frac{|\mathbf{K}\cdot\mathbf{W}|^2}{K^2} \right)
\, .
\label{eq:N_expect}
\end{equation}
The integrals on the right-hand side of equation \eqref{eq:N_expect} can be calculated exactly, and after some algebra we obtain
\begin{equation}
\langle \hat{N} \rangle = \frac{a}{64\pi^6}\,(n^2+m^2+l^2+s^2)
\, .
\label{eq:N_expect_1}
\end{equation}
As we can see from equations \eqref{eq:H_expect_1} and \eqref{eq:N_expect_1}, the energy and photon number are proportional to each other $\langle \, : H : \, \rangle = 2\hbar c \langle \, : N : \, \rangle$. This is consistent with each photon carrying average energy $2\hbar c$ in our units.
The ratio $\langle \, : \hat{H} : \, \rangle/\langle \, : \hat{N}\, : \rangle = 2\hbar c$ is independent of the integers and of $L_{0}$.  This is a physically sensible result, as each photon carries an average energy $2\hbar c$ which originates from the exponential cutoff $e^{-2k}$ (in $\mathbf{k}$ variables) that fixes a characteristic wave number $\sim 1$ in dimensionless units.

%Correlation functions
To conclude this section, let us discuss the coherence of 
$| \text{knot}\rangle$. We note that the positive-frequency part of the classical electric field is given by 
\begin{equation}
\mathbf{E}_{\mathrm{cl}}^{(+)}(\mathbf{x},t) = \frac{i\sqrt{a} L_0}{2^{3/2}\pi^2} \sum_{\lambda} \int \frac{d^3k}{(2\pi)^3} e^{-K} \bigl( \boldsymbol{\epsilon}_{\mathbf{k}\lambda}^* \!\cdot\! \mathbf{W}(\mathbf{K}) \bigr) \boldsymbol{\epsilon}_{\mathbf{k}\lambda} e^{i\mathbf{k}\cdot\mathbf{x} - i\omega_{\mathbf{k}}t}
\, ,
\label{eq:Eplus_integral}
\end{equation}
which is a deterministic function, as is the negative-frequency part $\mathbf{E}_{\mathrm{cl}}^{(-)}(\mathbf{x},t)$.  Therefore, the general definitions of the first- and second-order correlation functions from equations \eqref{eq:corr_fct_1} and \eqref{eq:corr_fct_2}, and the normalized correlations from equations \eqref{eq:hopfion_g1_final} and \eqref{eq:hopfion_g2_final} can be applied to the knotted coherent states. If one wishes to write $G^{(1)}$ and $G^{(2)}$ in terms of the parameter integrals, we can substitute the expression for $\mathbf{E}_{\mathrm{cl}}^{(+)}$ and obtain, for example, 
\begin{equation}
G^{(1)}(x,y) 
= 
\frac{a}{2^{5}\pi^4 L_0^4} \left( \int \frac{d^3K}{(2\pi)^3} e^{-K} \mathbf{W}^*(\mathbf{K}) e^{-i\frac{\mathbf{K}}{L_0}\cdot\mathbf{x} + i\frac{cK}{L_0}t_x} \right) 
\cdot 
\left( \int \frac{d^3K'}{(2\pi)^3} e^{-K'} \mathbf{W}(\mathbf{K}') e^{i\frac{\mathbf{K}'}{L_0}\cdot\mathbf{y} - i\frac{cK'}{L_0}t_y} \right)
\, .
\label{eq:G1_knot}
\end{equation}
This double integral is the explicit representation of $G^{(1)}(x,y)$, which is not very illuminating. 
However, we note that the corresponding correlation $g^{(1)}$ is not constant in general: its modulus equals $1$ only if the field is scalar or if the vector fields are parallel. 
For the knotted fields, the direction of $\mathbf{E}_{\mathrm{cl}}^{(+)}$ varies with position, so $|g^{(1)}(x,y)| < 1$ in general.   On the other hand, the normalized second-order correlation is given by \eqref{eq:hopfion_g2_final} as in the general case
\begin{equation}
g^{(2)}(x,y) = \frac{G^{(2)}(x,y)}{\langle \hat{I}(x)\rangle \langle \hat{I}(y)\rangle} = 1
\, ,
\label{eq:g2_knot}
\end{equation}
where $\hat{I}(x) = \hat{\mathbf{E}}^{(-)}(x)\cdot\hat{\mathbf{E}}^{(+)}(x)$ is the intensity operator. This general relation holds regardless of the specific form of $\mathbf{E}_{\mathrm{cl}}^{(+)}$, because $\langle \hat{I}(x) \rangle = |\mathbf{E}_{\mathrm{cl}}^{(+)}(x)|^2$. 
Thus, we can conclude that the knotted electromagnetic field is not fully coherent in the sense of equal polarization directions, but it is still a coherent state in the quantum optical sense, i.e., it is a minimum uncertainty state with Poisson statistics.

\subsection{Hopfion Coherent State}
\label{sec:examples}

As an example of the previous construction, we consider the Hopfion state defined by $n=m=l=s=1$. Its energy is given by formula \eqref{eq:H_expect_1} and takes the value
\begin{equation}
\langle :\!\hat{H}\!: \rangle 
= 2 \hbar c \langle :\! \hat{N} \! : \rangle 
= \frac{\hbar c a}{8\pi^6 }
\, .
\label{eq:Hopf_part_energy_number}
\end{equation}
From equation \eqref{eq:S_expect_1}, we get the Poynting vector 
\begin{equation}
\langle :\!\hat{\mathbf{S}} (\mathbf{X},T) \!:\rangle 
= \frac{8 a c }{\mu_0\pi^2 L_0^4 (1+R^2+T^2)^4}\,
\begin{pmatrix}
TX \\ TY \\ TZ - \frac{1+R^2-T^2}{2}
\end{pmatrix}
\, .
\label{eq:Hopf_S}
\end{equation}
Also, from equations \eqref{eq:Spin_expect} and \eqref{eq:Helicity_magnetic_comp}, we obtain the spin and magnetic helicity expectation values
\begin{equation}
\langle \, : \hat{\Lambda} : \, \rangle = \frac{a}{16\pi^6 L_0^2}
\, ,
\qquad
\langle \, : \hat{H}_{\mathrm{m}} \, : \rangle = \frac{a}{80\pi^6 L_0}
\, .
\label{eq:Hopf_Spin}
\end{equation}

The Hopfion coherent state discussed here is the simplest non-trivial topological electromagnetic configuration. As the above results show, its energy and photon number are proportional to $a/\pi^6$, showing that the parameter $a$ controls the quantum intensity. 
As in the general case, the Poynting vector exhibits a toroidal flow pattern: at $T=0$, it points along $Z$, and as time evolves, its components in the $X$ and $Y$ directions take non-zero values,  showing the energy circulation around the torus. 
The spin helicity $\langle \, : \hat{\Lambda} : \, \rangle$ is positive for our polarization convention. This can be interpreted as a net excess of right-handed circularly polarized photons. 
The magnetic helicity $\langle \, : \hat{H}_{\mathrm{m}} : \, \rangle$ is smaller by a factor $1/(5L_0)$ and shows the topological linking of magnetic field lines. 
This example explicitly shows that the coherent state inherits topological properties of the classical field, since the linking number $nm=1$ appears in the helicity expectations, and the spatial structure of observables follows the toroidal geometry.

\section{Conclusions}
\label{sec:conclusion}

In this paper, we have constructed explicit coherent state representations for the general Ra\~{n}ada-Hopf and non-null torus knot solutions of Maxwell's equations in vacuum. We used the general relation between classical electromagnetic fields and coherent state amplitudes to construct the general integral representations of coherent state amplitudes and displacement operators for Ra\~{n}ada-Hopf fields, as well as for their observables and correlation functions, and verified the defining properties of coherent states through direct computation.
For non-null torus knots, we computed exactly the vacuum expectation values of all physical observables of the system in terms of the integer parameters $(n,m,l,s)$ of the classical solutions. The analysis of correlation functions show that the topological information determines the quantum expectation values, while the state saturate the uncertainty relations and is Poissonian.

There are several future research directions that result from this work. One can generalize the present construction to squeezed states by applying the squeezing operator to the knotted coherent states. Studying the interplay between squeezing and topology could reveal new non-classical effects. Another line of research is based on the interaction of knotted coherent states with matter. As the results from literature on classical knotted fields show, this is expected to hold in the limit of weak charges (test particles), as the knotted states exists strictly in vacuum. Also, it is interesting to investigate the entanglement structure of multi-mode knotted coherent states, as well as their behaviour under decoherence, which could give information on the topological properties and their protection against certain noise channels. This might have implications for quantum information processing with light.

Other possible applications of these states could be speculated upon, like the precision measurements using the quantized helicity of topological light, the engineering of photonic devices with custom-tailored quantum correlations, and the simulation of topological field theories in table-top quantum optics experiments. However, further studies on the generation and interaction of these states with matter are necessary in order to derive experimental results.

\subsection*{Acknowledgments}

The authors thanks to C. M. Porto and C. F. L. Godinho for fruitful discussions. 
I. V. Vancea received support from the Basic Research Grant (APQ1)
from the Carlos Chagas Filho Foundation for Research Support of the State of Rio de Janeiro
(FAPERJ), grant number E-26/210.511/2024.

\end{document}